\documentclass[aps,prl,twocolumn,groupedaddress]{revtex4}
\usepackage[dvips]{graphics, color}

\usepackage{graphicx}
\usepackage{dcolumn} 
\usepackage{bm}

\begin{document}

\title{Probing the excitation spectrum of a Fermi gas in the BCS-BEC crossover regime}

\author{M. Greiner}
\email[Email: ]{markus.greiner@colorado.edu}
\homepage[URL:]{http://jilawww.colorado.edu/~jin/}
\author{C. A. Regal}
\author{D. S. Jin}
\thanks{Quantum Physics Division, National Institute of Standards and Technology.}
\affiliation{JILA, National Institute of Standards and Technology
and University of Colorado, and Department of Physics, University of
Colorado, Boulder, CO 80309-0440}

\date{\today}

\begin{abstract}
We measure excitation spectra of an ultracold gas of fermionic
$^{40}$K atoms in the BCS-BEC crossover regime. The measurements are
performed with a novel spectroscopy that employs a small modulation
of the B-field close to a Feshbach resonance to give rise to a
modulation of the interaction strength. With this method we observe
both a collective excitation as well as the dissociation of
fermionic atom pairs in the strongly interacting regime.  The
excitation spectra reveal the binding energy / excitation gap for
pairs in the crossover region.
 \end{abstract}


\maketitle

The predicted crossover between Bardeen-Cooper-Schrieffer (BCS) type
superfluidity and Bose-Einstein condensation (BEC) of molecules
\cite{Eagles1969a,Leggett1980,Nozieres1985,Randeria1995,Holland2001a,Timmermans2001a,Ohashi2002a}
has recently become experimentally accessible with a strongly
interacting gas of ultracold fermionic atoms
\cite{Greiner2003b,Jochim2003b,Zwierlein2003b,Bourdel2004a,Bartenstein2004a,
Regal2004a,Zwierlein2004a,HuletPrivate,Bartenstein2004b,Kinast2004a}.
The atom-atom interactions in such a gas, characterized by the
$s$-wave scattering length $a$, can be widely tuned with a
magnetic-field Feshbach resonance
\cite{Stwalley1976b,Tiesinga1993a,Inouye1998a}. On the BEC side of
the Feshbach resonance, the interactions are strong and repulsive
$(a>0)$, and there exists a weakly bound molecular state.
Bose-Einstein condensation of molecules in this state has been
observed
\cite{Greiner2003b,Jochim2003b,Zwierlein2003b,Bourdel2004a}. On the
BCS side of the resonance the gas has strong attractive interactions
$(a<0)$. While no two-body bound molecular state exists in this
region, many-body effects can give rise to pairing. Condensates of
these pairs have recently been observed in gases of $^{40}$K
\cite{Regal2004a} and $^6$Li atoms \cite{Zwierlein2004a}. The wide
Feshbach resonances used in these experiments arise from strong
coupling between open and closed scattering channels; therefore
these gases are expected to be well described by BCS-BEC crossover
physics \cite{Diener2004b} (see also \cite{Chen2004a} and references
therein).

To probe the many-body state and the nature of the pairs, the
excitation spectrum of the gas can be measured. Measurements of
collective excitations, for example, have provided further evidence
for superfluidity in the crossover regime
\cite{Kinast2004a,Bartenstein2004b}. To probe single (quasi-)
particle excitations, rf spectroscopy can be performed by applying a
radio-frequency (rf) field and inducing transitions to different
Zeeman or hyperfine levels. This rf spectroscopy has been used to
measure mean-field shifts and interaction strengths
\cite{Regal2003b, Gupta2003b}. In addition, on the BEC side of the
Feshbach resonance rf excitation was used to measure the
dissociation spectra and binding energies of weakly bound molecules
\cite{Regal2003c}. Recently, this technique has been applied to
measure the pairing gap in the BCS-BEC crossover regime
\cite{Baranov1999,Torma2000a,Chin2004a}.

Here we report on a new spectroscopy that takes advantage of the
tunability of interactions near a Feshbach resonance and probes the
excitation spectrum in the BCS-BEC crossover region. We excite the
system by modulating the interaction strength in the gas; this is
accomplished using a small sinusoidal modulation of the $B$-field
close to the Feshbach resonance. We find that the $B$-field
modulation can cause a dissociation of bound molecules or
generalized Cooper pairs into the free atom continuum. In contrast
to rf spectroscopy this method does not involve additional spin
states, but rather probes the excitation spectrum directly with no
frequency shifts due to the properties of additional spin states.
Also unlike rf spectroscopy, the $B$-field modulation does not drive
single-atom excitations.

\begin{figure}[h]
\begin{center}
\includegraphics[width=50mm]{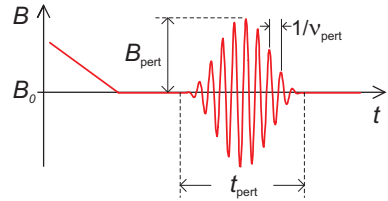}
\caption{Modulation sequence: After ramping to the final
magnetic-field value $B_0$, the magnetic field is sinusoidally
modulated at a frequency $\nu_\mathrm{pert}$. The envelope of the
modulation is a haversine function with a maximum amplitude
$B_\mathrm{pert}$ and a total duration
$t_\mathrm{pert}$.}\label{fig:modulation}
\end{center}
\end{figure}

The experimental setup and procedure are similar to that in our
previous work \cite{Regal2003b,DeMarco1999a}. We trap and cool a
dilute gas of the fermionic isotope $^{40}$K, which has a total spin
$f=9/2$ in its lowest hyperfine ground state and thus ten available
Zeeman spin states $|f,m_f\rangle$. The experiment is initiated by
preparing a nearly equal, incoherent mixture of the
$|9/2,-9/2\rangle$ and $|9/2,-7/2\rangle$ spin states. The atoms, as
well as the molecules we create from these atoms, are trapped in a
far off-resonant optical dipole trap. The trap is formed by a
gaussian laser beam at a wavelength of $\lambda=1064$\,nm focused to
a waist of 15.2\,$\mu$m. We evaporatively cool the atoms to
ultra-low temperatures by gradually decreasing the trap depth. Our
calculated final trap depth, including gravitational sag, is
$h\times9.6$\,kHz, where $h$ is Planck's constant. The measured
Fermi energy in this trap is $h\times4.6\pm0.6$\,kHz
\cite{TFcomment} in the weakly interacting regime. The calculated
radial trapping frequency in the center of the gaussian profile is
$\nu_r=$345\,Hz. We measure an effective radial trap frequency of
243\,Hz for the weakly interacting gas; this is consistent with the
trap frequency expected for a particle at about the Fermi energy in
the non-harmonic trapping potential. At the final trap depth the
optical trap holds $N_0=(1.8\,\pm0.9)\times10^5$ atoms per spin
state and the temperature of the weakly interacting gas, determined
by fits to the cloud in expansion, is $T/T_F=0.05\pm 0.02$.

The interaction between atoms in the two spin states is widely
tunable with an $s$-wave magnetic-field Feshbach resonance, which is
located at $202.10\pm0.07\,$G \cite{Regal2004a} and has a width of
$7.8\pm0.6$\,G \cite{Regal2003a}. Initially we prepare a quantum
degenerate Fermi gas in the weakly interacting region far on the BCS
side ($\Delta B>0$) of the Feshbach resonance. Then we reduce the
magnetic field adiabatically to bring the gas isentropically into
the region of strong attractive atom-atom interactions. For B-fields
up to 500\,mG away from the resonance on the BCS side we observe a
condensate of pairs of fermionic atoms \cite{Regal2004a}. These
pairs are attributable to many-body effects and can be regarded as
generalized Cooper pairs in the BCS-BEC crossover regime.

We measure the excitation spectrum of the gas at different final
magnetic field values $B_0$ by perturbatively modulating the
magnetic field around $B_0$ with a frequency $\nu_\mathrm{pert}$
(fig. \ref{fig:modulation}). This sinusoidal $B$-field modulation
causes a modulation of the effective atom-atom interaction strength.
Starting at $t=0$ we modulate the B-field for a total duration
$t_\mathrm{pert}$ between 5\,ms and 20\,ms. The modulation envelope
is given by a haversine function with a maximum amplitude
$B_\mathrm{pert}$ (see fig. \ref{fig:modulation}). In order to keep
the response perturbative, we scaled the applied modulation
amplitude as $B_\mathrm{pert}=b_0/\nu _\mathrm{pert}^{1/2}$, where
$b_0$ ranges between 75\,mG\,kHz$^{1/2}$ close to the resonance and
220\,mG\,kHz$^{1/2}$ far from the Feshbach resonance.

To quantify the response of the gas, we determine the increase in
the gas temperature after the perturbation is applied. We have found
that a sensitive way to detect this increase in temperature is to
measure the number of atoms that escape from the shallow trap due to
evaporation. The evaporated atoms are accelerated downwards due to
gravity and form a beam of atoms leaving the trap that can be
observed by resonant absorption imaging. By fitting the absorption
images to an empirical function we measure the number of atoms,
$N_\mathrm{evap}$, that are evaporated in the time between the start
of the modulation and 4\,ms after the modulation:

\begin{figure}[h]
\begin{center}
\includegraphics[width=86mm]{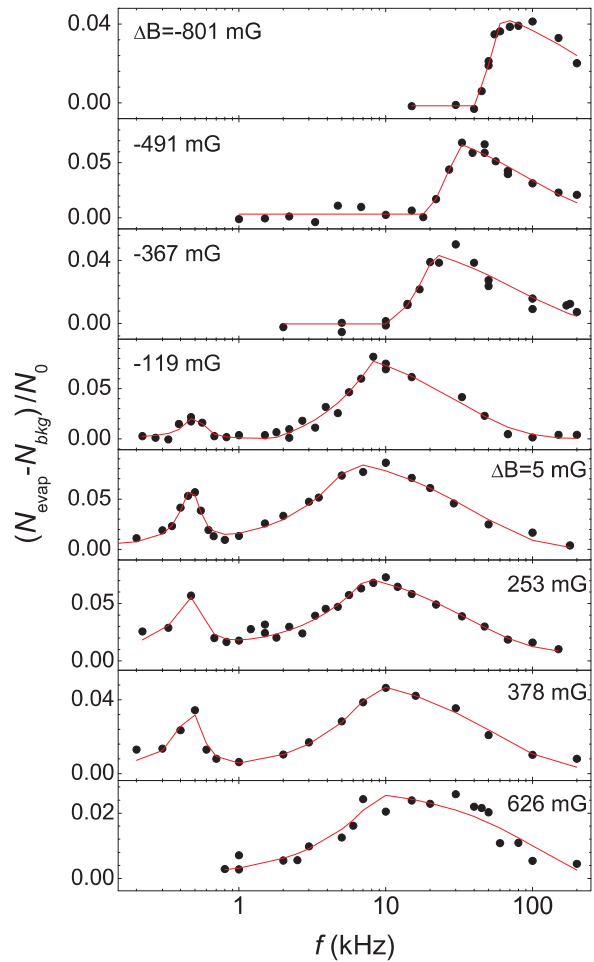}
\caption{Excitation spectra for different detunings $\Delta
B=B_0-202.10\,$G with respect to the Feshbach resonance. The
relative number of evaporated atoms $(N_\mathrm{evap}-N_{bkg})/N_0$
is a measure of the response and is plotted versus the modulation
frequency $\nu_\mathrm{pert}$. The modulation amplitude was scaled
as $B_\mathrm{pert}=b_0/\nu _\mathrm{pert}^{1/2}$ to keep the
response perturbative. The line is a fit to an empirical function,
from which we can extract the threshold position $\nu_0$ and the
position $\nu_{max}$ and height $\tilde{N}_{max}$ of the
maximum.}\label{img:NversNu}\end{center}
\end{figure}
\begin{equation}
N_\mathrm{evap}=\int_{0}^{t_\mathrm{pert}+4\,\mathrm{ms}}\dot{N}(t)dt.
\end{equation}

Figure \ref{img:NversNu} shows measured excitation spectra for
different interparticle interactions. The relative number of
evaporated atoms $(N_\mathrm{evap}-N_{bkg})/N_0$, which provides a
measure of the excitation response, is plotted versus the frequency
of the perturbation $\nu_\mathrm{pert}$ on a logarithmic scale.
Here, $N_{bkg}$ is the small number of evaporated atoms observed
with no applied perturbation. The following features can be seen:
\begin{itemize}

\item We find a distinct peak in the excitation spectra for modulation
frequencies $\nu_\mathrm{pert}\approx500$\,Hz, which is close to
twice the trap frequency. We attribute this peak to a collective
excitation of the trapped gas driven by the periodic modulation of
the interaction strength.
\item For frequencies larger than a threshold $\nu_0$ the
response increases. We interpret this threshold as a dissociation
threshold: Pairs are only dissociated if $h\nu_\mathrm{pert}$ is
larger than the effective pair binding energy. On the BEC side of
the resonance $\nu_0$ is nonzero and increases for decreasing $B_0$,
consistent with two-body predictions for molecule binding energies.
\item For increasing frequency beyond $\nu_0$ the response reaches
a maximum and then slowly decreases.
\end{itemize}

We have measured the dependence of the dissociation response of the
gas on the amplitude and duration of the B-field modulation. The
number of evaporated atoms depends linearly on the duration and
quadratically on the amplitude of the modulation, with a small
offset due to a constant background heating and evaporation rate.
The response due to the dissociation of pairs can be understood as
follows: After dissociation the two free atoms gain a total relative
kinetic energy $\Delta E=h\nu_\mathrm{pert}-E_b$, where $E_b$ is the
effective pair binding-energy or twice the excitation gap and $h$ is
Planck's constant. Since the sample is collisionally dense in the
region of strong interactions close to the Feshbach resonance, the
dissociated atoms undergo multiple collisions and deposit this
energy into the gas. We have verified that the measured response,
$(N_\mathrm{evap}-N_{bkg})/N_0$, is proportional to the heating
measured directly in a deeper trap. The spectra in fig.
\ref{img:NversNu} are then proportional to
\begin{equation}
\Gamma(\nu)\frac{h\nu_\mathrm{pert}-E_b}{h\nu_\mathrm{pert}},
\label{eqn:heating}
\end{equation}
where $\Gamma(\nu)$ is the frequency dependent dissociation rate and
the factor $(h\nu_\mathrm{pert})^{-1}$accounts for the modulation
amplitude being scaled as $(h\nu_\mathrm{pert})^{-1/2}$.

For a quantitative study we have fit the data to an empirical model
(line in fig. \ref{img:NversNu}). The collective excitation peak at
low modulation frequencies is fit with a Lorentz curve. The
subsequent increase of the response is fit to a linear slope,
starting at the threshold $\nu_0$ (note that the linear curve
appears bent in the logarithmic plot). After the maximum, the fit
function consists of an exponential decay.

\begin{figure}[h]
\begin{center}
\includegraphics[width=66mm]{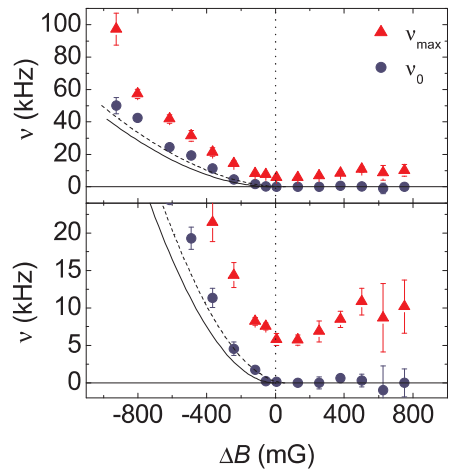}
\caption{Threshold frequency $\nu_0$ (circles) and frequency of
maximum response to perturbation $\nu_{max}$ (triangles) versus
detuning from Feshbach resonance $\Delta B$. The lower graph is a
vertical zoom of the upper graph. For comparison, the two-body
calculation of the molecule binding energy is shown for a resonance
centered at $\Delta B=0$ (solid line) and for a resonance position
shifted to $\Delta B=70$\,mG (dashed line). }\label{img:threshold}
\end{center}
\end{figure}

\begin{figure}[h]
\begin{center}
\includegraphics[width=75mm]{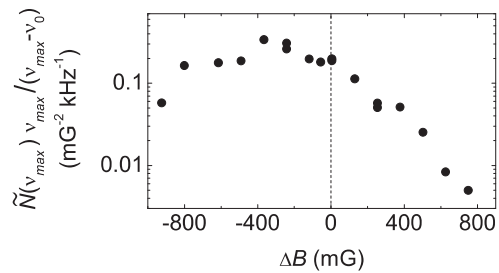}
\caption{Dissociation response for various detunings $\Delta B$,
plotted as the maximum value of the scaled number of evaporated
atoms $\tilde{N}(\nu_{max})$ with the scaling factor
$\nu_{max}/(\nu_{max}-\nu_0)$ (see text). }\label{img:amplitude}
\end{center}
\end{figure}

Figure \ref{img:threshold} shows the positions of the fitted
threshold $\nu_0$ and the response maximum $\nu_{max}$ versus the
B-field relative to the Feshbach resonance position,  $\Delta B$.
For comparison we also show a plot of the two-body binding energy in
units of Hertz from a full coupled channel calculation
\cite{ChrisPrivate} on the BEC side of the Feshbach resonance. The
measured threshold for dissociation is close to the two-body binding
energy of the molecules. The data might suggest that the actual
position of the Feshbach resonance is at slightly higher magnetic
fields, closer to the upper limit (+70\,mG) given in ref.
\cite{Regal2004a} (dotted line). However, the data in fig.
\ref{img:threshold} are measured at large densities and strong
interaction strengths where mean-field shifts, arising from the
differences between molecule and atom interactions, are expected to
play a significant role.

On the BCS side we find that the threshold, $\nu_0$, is close to
zero, suggesting that some atom pairs can be broken by arbitrarily
small perturbation energies. We believe that this is a consequence
of the density inhomogeneity of the trapped gas. The pairing gap
arises from many-body effects and should decrease in the low density
regions away from the center of the trap. The frequency location of
the maximum signal then provides information about the density
dependent pairing gap. We find that the shape of the excitation
spectrum is approximately constant on the BCS side and the maximum
of the signal occurs at $\nu_{max}$ between 6\,kHz and 11\,kHz,
which is close to twice the Fermi energy. In our experiment we see
no significant qualitative change in the spectra when the
experiments are performed at slightly higher temperatures above the
observed pair condensation temperature. This is consistent with the
expectation that atoms pair up with a pseudogap prior to
condensation with a superfluid gap \cite{Chen2004a}. A quantitative
theory of the system and its excitations might allow one to extract
further information about the pseudogap and/or superfluid gap from
the data.

In fig. \ref{img:amplitude} we plot a measure of the response
strength, $\tilde{N}(\nu_{max})\,\nu_{max}/(\nu_{max}-\nu_0)$,
versus the $B$-field $\Delta B$, where
$\tilde{N}=(N_\mathrm{evap}-N_{bkg})/(b_0^2\,t_\mathrm{pert})$ is
the scaled number of evaporated atoms \cite{note:tpertCorrection}.
Here, we divide $\tilde{N}(\nu_{max})$ by $(\nu_{max}-\nu_0)$ to
account for the heating being proportional to the excess energy
above threshold and multiply by $\nu_{max}$ to account for our
frequency dependent modulation amplitude (see eq.
\ref{eqn:heating}). We find a strong perturbation signal close to
the Feshbach resonance on both sides. The signal greatly decreases
for larger positive detunings from the resonance. However, it is
still nonzero in a region $\Delta B>500$ mG where the measured
condensate fraction vanishes. This suggests that in this region
non-condensed pairs are present, which would be consistent with a
pseudogap theory \cite{Chen2004a}.

In a simple two-body model, in which the interatomic potential is
described by a square well and molecules on the BEC-side of the
Feshbach resonance are dissociated by periodically modulating the
potential depth, we calculate the dissociation rate,
$\Gamma(\nu_{pert})$, to be proportional to
$\sqrt{h\nu_\mathrm{pert}-E_b}/h\nu_\mathrm{pert}$. However,
empirically we find the BEC-side spectra to be more consistent with
a rate that goes as
$\sqrt{h\nu_\mathrm{pert}-E_b}/(h\nu_\mathrm{pert})^2$. This
suggests that a more sophisticated model is needed.

In conclusion we have introduced a new method for measuring
excitation spectra in the BCS-BEC crossover regime. In addition to
driving a collective oscillation we find that modulating the
interaction strength, by modulating the magnetic field, dissociates
fermionic atom pairs. Excitation spectra have been recorded for
various interaction strengths, and dissociation thresholds and
maxima have been determined. This novel spectroscopy can provide a
probe of pseudogap physics and superfluidity at the BCS-BEC
crossover.

We thank M. L. Chiofalo, C. H. Greene, C. Ticknor and J. L. Bohn for
stimulating discussions. This work was supported by NSF and NIST. C.
A. R. acknowledges support from the Hertz Foundation.

\bibliographystyle{prsty}

\begin{thebibliography}{10}
\bibitem{TFcomment}
Here $T_F$ is determined from a Thomas Fermi fit. If $T_F$ is
instead
  determined based on the momentum spread, the measured trap frequencies and
  atom number, we get $T_F=h\times6.0\,\pm0.6$\,kHz, where the uncertainty is
  statistical. This is in reasonable agreement given the uncertainty in atom
  number and trap frequency.
\bibitem{note:tpertCorrection}
For $t_\mathrm{pert}\geq10$\,ms we apply a correction factor
determined
  experimentally from our measurement of the response versus the perturbation
  time.
\end{thebibliography}

\end{document}